# Longitudinal integration measure in classical spin space and its application to first-principle based simulations of ferromagnetic metals.

Sergii Khmelevskyi

Center for Computational Materials Science, Vienna University of Technology, Wiedner Hauptstrasse 8-10, A-1040, Vienna, Austria.

The classical Heisenberg type spin Hamiltonian is widely used for simulations of finite temperature properties of magnetic metals often using parameters derived from first principles calculations. In itinerant electron systems, however, the atomic magnetic moments vary their magnitude with temperature and the spin Hamiltonian should thus be extended to incorporate the effects of longitudinal spin fluctuations (LSF). Although the simple phenomenological spin Hamiltonians describing LSF can be efficiently parameterized in the framework of the constrained Local Spin Density Approximation (LSDA) and its extensions, the fundamental problem concerning the integration in classical spin space remains. It is generally unknown how to integrate over the spin magnitude. Two intuitive choices of integration measure have been used up to date - the Murata-Doniach scalar measure and the simple three dimensional vector measure. Here we derive the integration measure by considering a classical limit of the quantum Heisenberg spin Hamiltonian under conditions leading to the proper classical limit of the commutation relations for all values of the classical spin magnitude and calculate the corresponding ratio of the number of quantum states. We show that the number of quantum states corresponding to the considered classical spin magnitude is proportional to this magnitude and thus a non-trivial integration measure must be used. We apply our results to the first-principles simulation of the Curie temperatures of the two canonical ferromagnets bcc Fe and fcc

Ni using a single-site LSF Hamiltonian with parameters calculated in the LSDA framework in the Disordered Local Moment approximation and a fixed spin moment constraint. In the same framework we compare our results with those obtained from the scalar and vector measures.

The great success of ab-initio methods in solid state physics and materials science is based on the wide use of the Density Functional Theory[1] (DFT) and the Local (Spin) Density Approximation (L(S)DA)[2], which provides a convenient tool for the estimation of the ground state energy of an extended electron system in a mean-field framework. Varieties of the L(S)DA extensions including an approximate treatment of correlation effects[3,4] allow a robust description of the ground state properties for a wide range of materials. However an ab-initio based description of finite temperature effects, in particular of entropy driven phase transitions is often based on appropriately chosen model approaches and first-principles calculations are used for the estimation of the model parameters.

Perhaps one of the most important and transparent cases is the first-principles modeling of the finite temperature magnetism in solids. The basic phenomenological model applied is the Heisenberg model describing the interaction of localized atomic spins on the lattice:

$$H_{quantum} = -\sum_{i,j} J_{ij} \hat{S}_i \hat{S}_j, \qquad (1)$$

where $\hat{S}_i$ is the quantum spin-operator acting on the lattice site $i$ and $J_{ij}$ are interatomic exchange constants[5]. The quantum Heisenberg Hamiltonian (1) and its extensions are the backbone for modern theory of the magnetism for systems with localized atomic moments. In practice, however, estimations of the magnetic Hamiltonian parameters from first-principles calculations[6] are performed for the classical spin Hamiltonian, which can be derived from (1) by taking the classical limit ($\hbar \to 0$, $S \to \infty$), where S is spin moment at the given atomic site:

$$H_{cl.} = -\sum_{ij} J_{ij} \vec{m}_i \vec{m}_j, \qquad (2),$$

where $\vec{m}_i$ is classical spin vector representing the spin moment of the atom on site $i$. The length of the vector $\vec{m}_i$ we will call the spin moment magnitude and it is equal (in atomic units) to $\sqrt{S(S+1)}$. In the following we will also deal with the spin models where magnitude m can takes continuous set of the values.

There are few effective ways of ab-initio estimation of the exchange constants based either on a direct mapping of the total energies calculated for various magnetic configurations onto the Hamiltonian (2), (e.g. from non-collinear spin-spiral calculations[7]) or the use of the Green function based magnetic force theorem, formulated in an ab-initio framework by Lichtenstein *et al.*[8] and its relativistic generalization[9]. The number of applications of this scheme has been growing almost exponentially during the last two decades. Nowadays, the application of the Hamiltonian (2) with ab-initio calculated parameters has become an almost routine procedure[6] and when used in atomistic spin-dynamics simulations it constitutes a first step in the multi-scale magnetic modeling paradigm.[10,11] The use of the classical Hamiltonian (2) instead of the quantum formulation (1) is not just an artifact of the LSDA approximation, since the case of metallic (itinerant) magnetism it rather a consequence of the band character, itinerancy, of the d-electrons forming atomic magnetic moment.[12] For the further discussion of this problem we can refer to the works of Körmann *et al.*[13,14] In addition there is a wide use of classical spin models in atomistic spin dynamics simulations.[15,16] In this manuscript we consider one of the most peculiar problem related to the application of the Hamiltonian (2) for simulation of the finite temperature properties of the metallic systems with itinerant magnetic moments, where the magnitude of the magnetic moment on the atomic sites might change with temperature.

It has been realized earlier that the simulations done with the Hamiltonian (2) and the calculated exchange interactions might be successful only in case of well localized magnetic moments. For instance, the whole approach fails for Ni, since its atomic magnetic moment magnitude strongly depends on the temperature and the magnetic state.[17] More generally, in metallic transition metal system there are thermal excitations which lead to a change of the local moment magnitude. In the case of weak itinerant magnets these longitudinal spin fluctuations (LSF) are the leading mechanism responsible for the formation of the atomic moments in the high temperature phase[12,18]. However, even in the latter case the LSF contribution remains important.[18,19] The more a system deviates from local moment behavior the less effective first-principle based finite temperature simulation schemes based on (1) and (2) are. The problem of LSF effects has been challenged since the 70ies and appears to be rather complex. It has been proposed[12] that the best phenomenological description can be achieved by using a Hamiltonian, which gives a crossover between the local moment picture and the weak itinerant limit. Such a Hamiltonian must involve next to the inter-site Heisenberg-like interactions additional single-site terms, which describe the dependence of the energy on the local moment magnitude on the given site. Attempts to include the LSF effects into the ab-initio modeling are usually based on some form of such Hamiltonian. A phenomenological LSF Hamiltonian can be written as[12,18,19]

$$H_{LSF} = \sum_i E_i(|\vec{m}_i|) - \sum_{i,j} J_{ij} \vec{m}_i \vec{m}_j, \quad (3)$$

where the first term is the single-site energy which depends on the local moment magnitude.

The one commonly used method to map ab-initio LSDA results onto the Hamiltonian (3) is based on using spin-spiral total energy calculations.[7] This approach also requires using the fixed spin moment (FSM) constraint to explore the full configuration space.[20] The development and application of this method to LSF problem has been done in number of works.[21,22,23] An alternative approach is based on the LSDA based disordered local moment (DLM) approximation.[24,25,26] For the high temperature paramagnetic regime the on-site energies can be approximated as energies of the impurities in the DLM host[27] or in a slightly less accurate but far less resource demanding approximation, by the DLM energies itself.[28,29] One of the advantages using the DLM approximation is a possibility to simulate the exchange constants in the high temperature paramagnetic state rather than in some magnetically ordered reference state. For metallic systems, even in case of strong magnetic moment localization, these DLM exchanges might be quite a different than those derived from the ordered state, which applies even for the perfectly localized 4f-moments in hcp Gd metal[30,31]. Major steps towards the inclusion of the effects of itinerancy and LSF in first-principle based atomistic spin dynamic simulations were done recently by Antropov[32] and Ma and Dudarev.[33] The comparison of using the ferromagnetic and the DLM reference states in connection with a LSF Hamiltonian for magnetic alloy systems has recently been discussed.[34]

The overall success of LSF inclusion in ab-initio simulations of finite temperature magnetic is rather remarkable.[22,27,29,34] However, there is one fundamental problem related to the statistical simulations with the classical LSF Hamiltonian (3), namely that the exact form of measure for integration over the spin moment magnitude in classical spin configuration space is unknown, as pointed out by Wysocki *et al.*[18] Moreover, since the spin operator has no classical analogue, it cannot be derived from the correspondence principle.[18]

All previous ab-initio based approaches to the LSF problem used one or the other type of measure, even if it was not mentioned explicitly in the respective works. Most commonly, it was assumed that number of the spin states per volume of the phase space is distributed equally, for the spherically symmetric paramagnetic case without short range order this results in an integration measure $d\vec{m} = 4\pi m^2 dm$, where m is the atomic moment magnitude. This measure has been widely used[35] in spin-spiral based studies[21,22,23] and in some DLM based as well.[28,29] In the following we will denote this as the 3D vector measure (3D). In other studies[27,34] it has been assumed that the number of states corresponding to different spin magnitudes are equal leading to $d\vec{m} = 4\pi dm$. This metric has been introduced in the classical work by Murata and Doniach (MD).[36] To average over the spin magnitude in the high temperature paramagnetic state in mean-field approximation for the LSF Hamiltonian (3) one can write[23,28,29,34]

$$\langle m \rangle_T = \frac{\int_0^\infty m g(m) Exp\left(-E(m)/k_b T\right) dm}{\int_0^\infty g(m) Exp\left(-E(m)/k_b T\right) dm}, \quad (4)$$

where the measure $g(m)=1$ (MD) or $g(m)=m^2$ (3D) and $E(m)$ is the single-site term from the Hamiltonian (3). Up to now only applications and comparison with experiment have guided the choice of measure.[18] However, the problem of measure is much more important for reconciliation of various ab-initio approaches to LSF than possible shortcomings of LSDA in describing the relevant energetic since LDA can be extended to include the missing effects of electronic correlation by e.g. using LDA+U, self-interaction corrections, or LDA+DMFT. However, recent LDA+DMFT calculations of exchange constants for Fe and Ni suggest a good agreement with LSDA.[37] The problem of deriving a proper expression for the measure in classical phase spin space is about to provide arguments, which allow to estimating the relative weight of states having a different atomic classical spin (magnetic moment) magnitude. The estimation of this ratio is also needed for the solution of another long standing problem related to the physics of magnetic transition metals alloys. Already very early it has been realized that the magnetic entropy is an essential contribution to the structural and phase stability.[38] The ab-initio estimation of the magnetic entropy for metallic magnetic alloys, in particular for the most important case of Fe-based alloys, is a major problem in simulations. Today a simple generalization of the quantum mechanical entropy, $H = -k_b \ln(2S+1)$, where the, $S$ is value of the unpaired spin per magnetic atom, has been proposed as a solution.[38,39] Despite the fact that it is hard to justify this expression with non-integer $m$ values of spin moment, e.g. for the majority of magnetic transition metal alloys, the use of such an entropy term has become rather common in the CALPHAD community.[40] The formulation of the entropy in the form $k_b \ln(m+1)$ in the ab-initio framework has been advocated by Heine and Joynt,[41] but the widespread use of this idea in the framework of LSDA started just recently.[42,43,44,45,46,47,48] In this approach the free energy per magnetic atom can be written as a function of the local moment magnitude

$$f(m) = E_{DLM}(m) - TH = E_{DLM}(m) - k_b T \ln(m+1) \quad (5),$$

where $E_{DLM}$ is an magnetic energy, calculated from DLM-LSDA with FSM constraint (in real alloy simulations this energy might also include all other non-magnetic contributions, electronic energies, vibrations, configurational entropy etc.[44] Minimizing this expression with respect to $m$, one can calculate the value of the local moment magnitude at a given temperature. Justification of using magnetic entropy in the form $k_b \ln(m+1)$ or finding eventually more accurate expressions again can be linked to the observation of a relative number of states corresponding to two different values of the non-integer atomic moment magnitude since

$$\Delta H = H(m_1) - H(m_2) = k_b \ln(N(m_1)/N(m_2)) \quad (6),$$

where *N(m)* are the respective number of states. Ruban *et al.*[28] attempted to reconcile both approaches arguing that using the 3D measure with the LSF Hamiltonian corresponds approximately to a magnetic entropy of the form $k_b\, 3ln(m)$ in Equ.(5).

In order to derive a proper measure for the integration in the classical spin space one need to solve a full statistical many body problem for metal. Since the problem till now cannot be attack in this general form, in this work we provide simple arguments toward to the procedure of the measure construction by taking the proper classical limit of the quantum Heisenberg Hamiltonian (1) to the manifold of classical Hamiltonians (2) with various classical spin magnitudes. We must note that that quantum Hamiltonian (1) is used here entirely for estimation of the minimal phase volume in general classical configuration spin space with variable moment magnitude and has nothing to do with original quantum problem of electronic structure of the metals. For this purpose we recall the proper limiting procedure described by Månson,[49] who demonstrated that the right way to derive a classical Heisenberg Hamiltonian (2) from its quantum form (1) is to take the limits ($\hbar \to 0$, $S \to \infty$) subject to the additional condition that $\hbar \sqrt{S(S+1)} \to m$. Only if this additional condition is satisfied the quantum commutation relations for spin operators transform to the classical expressions for spin moment vectors and quantum equations of motions for the $\widehat{S}$ operators transform to those for classical angular momenta.[44] The idea of our work is to find a limit of the ratio $N(m_1)/N(m_2)$ by exploring the additional Månson condition and taking the classical limits of quantum Heisenberg Hamiltonians with different values of quantum number S. The number of quantum states which corresponds to the given *S* value is *2S+1* and it obviously tends to infinity in classical limit. However, their ratio for two limiting procedures tends to the finite value, which just a ratio of the corresponding classical magnitudess:

$$\frac{\lim\limits_{\hbar\to 0, S\to\infty:\sqrt{S(S+1)}\to m_1/\hbar}(2S+1)}{\lim\limits_{\hbar\to 0, S\to\infty:\sqrt{S(S+1)}\to m_2/\hbar}(2S+1)} = \lim_{\hbar\to 0}\frac{\lim\limits_{\sqrt{S(S+1)}\to m_1/\hbar}(2S+1)}{\lim\limits_{\sqrt{S(S+1)}\to m_2/\hbar}(2S+1)} = \lim_{\hbar\to 0}\frac{\sqrt{1+4m_1^2/\hbar^2}}{\sqrt{1+4m_2^2/\hbar^2}} = \frac{m_1}{m_2} \quad (7),$$

where we use Manson additional condition and reduce conditional double limit ($\hbar \to 0$, $S \to \infty$) to iterated one with free $\hbar \to 0$ external limit, and $m_1$ and $m_2$ are the classical moment magnitudes corresponding to two different values of the quantum number S. Thus, it appears that proper classical limit of quantum Heisenberg Hamiltonian dictates the relative number of states with different classical spin magnitude to be proportional to the ratio of respective magnitudes. The derivation is, strictly speaking, limited to discrete sets of magnitude values m corresponding to discrete set of quantum numbers S. However, one can immediately observe that neither of the "traditional" metrics g=1 or g=$m^2$, used for models with continuous magnitude, are compatible with equ.(6). Only metrics g=m provides the same result as the equ.(6) for a classical model with continuous m.

In the following we use the measure derived ($g(m) = m$) in conjunction with our LSF Hamiltonian (3) and the new expression for the entropy for an ab-initio estimation of the Curie temperature and the temperature dependent magnetic moment magnitude for fcc Ni and bcc Fe. Within the same computational framework we compare the performance of various forms of measures and entropies discussed above. To this end we apply the LSDA[50] and the Korringa-Kohn-Rostoker (KKR) method[51] in the same setup as it was used in previous works on LSF Hamiltonian mapping.[28,29,34] The exchange constants of the Hamiltonian (3) were calculated using the Green Function based magnetic force theorem8 as implemented in the KKR band structure code.[52]

We calculate the dependence of the energy of the paramagnetic DLM state[24] on the value of the local atomic moment magnitude for fcc Ni and bcc Fe at the experimental lattice constants. The results are shown in figure 1a. Nickel does not possess stable local moment in the DLM state whereas iron is relatively well localized, thus these metals represent the limiting cases of (itinerant) band-magnetic systems. The vertical bar in figure 1a illustrates the energy range of the LSF excitations at the experimental Curie temperatures. For Fe the local moment magnitude is not expected to change significantly independent on the choice of the measure and is expected to remain close to the self-consistently calculated DLM value (minimum of the curve). One sees that for Ni (and also for intermediate systems[29]) the choice of measure might be crucial, whereas for Fe Hamiltonian (2) is already a good approximation. Using equ.(4) and the DLM energies from Fig.1a we calculate the temperature dependence of the local moment magnitude by performing a numerical integration with different measures $g(m)$ – (Fig.1b for Fe) and (Fig.1(c) for Ni). It can be seen immediately that even for Fe the choice of the measure leads to a qualitatively different behavior of $m(T)$. The 3D measure leads to an increase of the spin moment whereas the MD measure to a decrease. The measure $g(m)=m$ on the other hand provides an almost temperature independent behavior. This partially explains, why various earlier ab-initio approaches for the finite temperature magnetism of bcc Fe, which completely ignore LSF, were so surprisingly successful.[17,53] It is also interesting to compare the paramagnetic moments of fcc Ni estimated approximately for near above the $T_c$ from neutron diffraction experiments (~0.4 $\mu_B$/Ni)[54] and calculated one with different measures. It is seen that the new measure provides rather good estimate compare to the other approaches.

Following the procedure described in Ref.[29]. we calculate the inter-atomic exchange constants in the DLM state with fixed local moment magnitudes. The temperature dependence of the average local moment magnitude in $V_3Al$, is calculated by the numerical integration (equ. 4) with use of the calculated DLM energies (Fig.1a). The result of this integration with three different metrics are presented in the Fig. 1b and Fig. 1c for Fe and Ni respectively.. We emphasize again that the procedure is justified only for the paramagnetic state above the magnetic ordering temperature. Then we perform Monte-Carlo (MC) simulations with the Hamiltonian (2) to derive a magnetic phase transition temperature corresponding to a set of fixed magnetic moment magnitude values. The real Curie temperature ($T_c$) is than estimated as the crossing point of m(T) and the M-C $T_c$(m) (see Fig.1). For bcc Fe both MD and $g(m)=m$ measures give a very good estimate close to the experimental[49] $T_c$ of 1043K, namely 990K and 1130K , respectively . In contrast, the 3D measure slightly overshoots value (1250K).

For bcc Ni, however, only the here derived measure $g(m)=m$ leads to a reasonable result (700K) compared to the experimental value[55] (627K). The 3D measure strongly overestimates $T_c$ (1250K) and MD measure strongly underestimates it (~200K). Note, that the strong underestimation of $T_c$ for fcc Ni with the MD measure[34] and the considerable overestimation of $T_c$ with the 3D measure for bcc Fe are both known features. It thus appears that the measure derived here from the limiting procedure, might be a reconciliation threshold for future applications of LSF theory in ab-initio modeling.

Let us now turn to the model based on the free energy given by Equ.(5). Here we must state clearly that the use of the procedure based on the equation (5) below is made just for illustrative purposes, since the approach based on the LSF Hamiltonian (3) presented above is more general. The information concerning the magnetic free energy of the paramagnetic state (and thus of the entropy) can be derived from the denominator of the equ. (4), which is just a partition sum. However we believe that the results of the use of the free energy expression in the form of equ. (5) with various forms of the entropy term and their comparison might be interesting to some readers due to existing number of works in the literature, which used the expression (5) as staring point (see e.g very recent work[48]). One need to mention, however, that the form of the entropy term might be different if one start from the LSF Hamiltonian (3) and making some assumptions concerning the analytical form of energy vs. moment dependence in way similar to presented in Ref. [28]. In this case the metrics g(m) = m might lead to the entropy proportional to the $k_b\, 2ln(m)$ rather than $k_b\, ln(m)$ [56].

We use the DLM energies from Fig.1a and three different expression for the magnetic entropy: $k_b\, ln(m+1)$, $k_b\, ln(m)$, and $k_b\, 3ln(m)$. Then we numerically minimize the free energy (Equ.5) with respect of m on a fine grid of temperature values. The resulting curves are given in figure 2. The estimation of $T_c$ is done with same procedure as described above – exchange constant mapping and Monte-Carlo simulations with the Hamiltonian (2). One notices immediately that for bcc Fe the use of Grimvall ($k_b\, ln(m+1)$) and $ln(m)$ entropies leads to the very similar results, whereas for fcc Ni the Grimvall entropy does not even predict a magnetic order. Using the expression ln(m) for the entropy for fcc Ni, we also find good agreement of the calculated $T_c$ (680K) with experiment (628K). Ruban's entropy ($k_b\, 3ln(m)$) overestimates $T_c$ in both cases, which is not surprising as it is derived as approximation to the results of LSF theory within the 3D measure. Both entropies $k_b\, ln(m+1)$ and $k_b\, ln(m)$ overestimate $T_c$ of Fe, however, to a much less extent compared to $k_b\, 3ln(m)$. We can thus arrive to an important conclusion: the limiting procedure described in this work justifies the use of a simple $k_b\, ln(m+1)$ entropy with non-integer magnetic moments, but only for well localized itinerant systems like bcc Fe. The more the system deviates from a local moment behavior (fcc Ni) the less reasonable results can be obtained with $k_b\, ln(m+1)$. Caveat: the whole approach described here can only be applied at high temperatures and in the paramagnetic regime. Moreover, one should notice that the computational approach we used here ignore the contribution to the entropy from lattice vibrations, major part of the electronic entropy (we deal only with electronic entropy differences related to the magnetic excitations) and thermal expansion effects which may change[57] the electronic structure and value of the exchange constants influencing the Curie temperature value.

We derive an expression for the integration measure in configuration space of the extended classical Heisenberg Hamiltonian with longitudinal variations of the spin moment magnitude which is consistent with a Manson conditions for classical limit of the Heisenberg Hamiltonian. This measure appears to be rather different from the two commonly used measures in present first-principle based simulations with phenomenological Hamiltonian, which include the variable atomic moment magnitude. Our application of this measure for the ab-initio estimation of the Curie temperature for two canonical metallic ferromagnetic metals, Fe and Ni, suggests that it provides the perspective to serve as a starting point for the reconciliation of various first-principles models of LSF in metals and alloys. This measure can be also used for first-principles based atomistic spin-dynamic simulations. In particular, since the Månson condition, used in our derivation, provides the correct classical limit for the dynamical equation of motion of the quantum spin operator[49]. The magnetic entropy in the form derived in this work, may be used as valuable alternative to the commonly used entropy $k_b\ ln(2S+1)$, which is only justified in the case of metals with integer spin per atom. The use of the longitudinal integration measure derived in this work can potentially provide a unifying platform for future simulations of longitudinal spin fluctuations effects in metallic magnets.

The author is grateful to Peter Mohn for useful discussion and critical reading of the manuscript. This work has been supported by FWF ViCom project F4109-N28.

**Figure captions.**

**Figure 1.(a)** Calculated total energies of the DLM states as function of the local spin moment magnitude (LSMA). Open symbols – fcc Ni, closed symbols - bcc Fe . The energies are given in Kelvin per magnetic site. The temperature dependence of the average LSMA in the paramagnetic state of Fe (panel b) and Ni (panel c). The values are derived by numerical integration of the DLM energies obtained with LSDA at each temperature point for the three different measures described in the text. Curie temperatures calculated for a set of LSMA values by Monte-Carlo simulations (open symbols) as described in the text. Experimental $T_c$'s are shown by vertical lines. (color online)

**Figure 2.** The result of the minimization of the free energy Equ.(5) calculated from the DLM energies from figure 1a, with various choices of the magnetic entropy expressions as described in the text. a) for bcc Fe, b) for fcc Ni. The lines with open symbols are Monte-Carlo simulations of $T_c$. (color online)

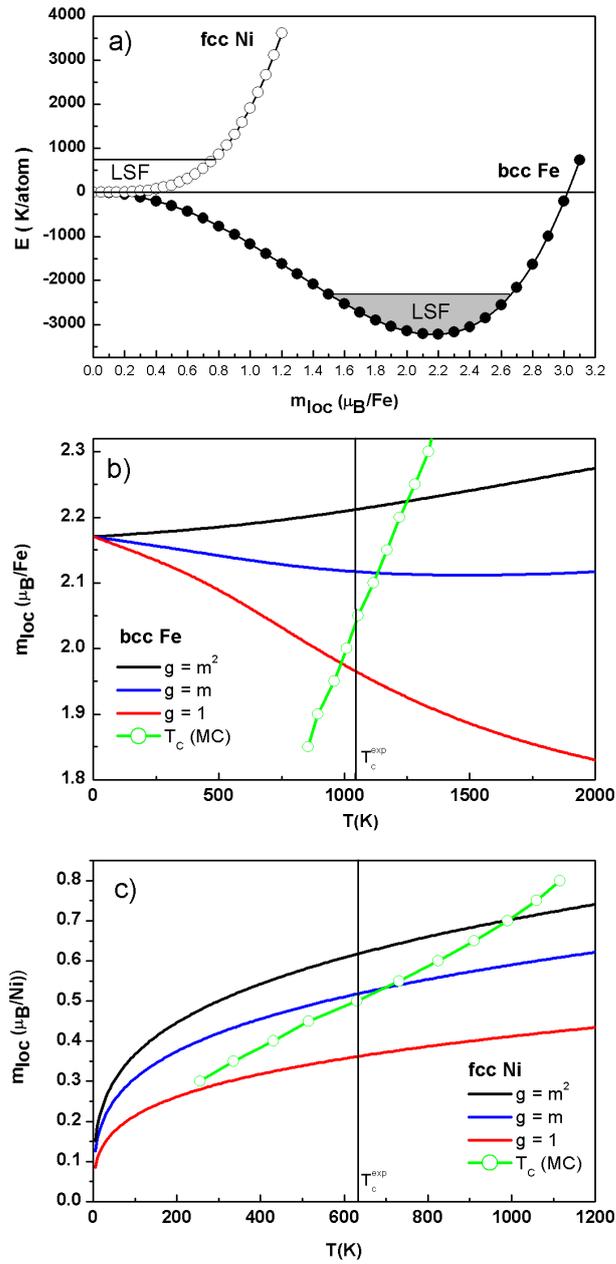

**Figure 1_Khmelevskyi**

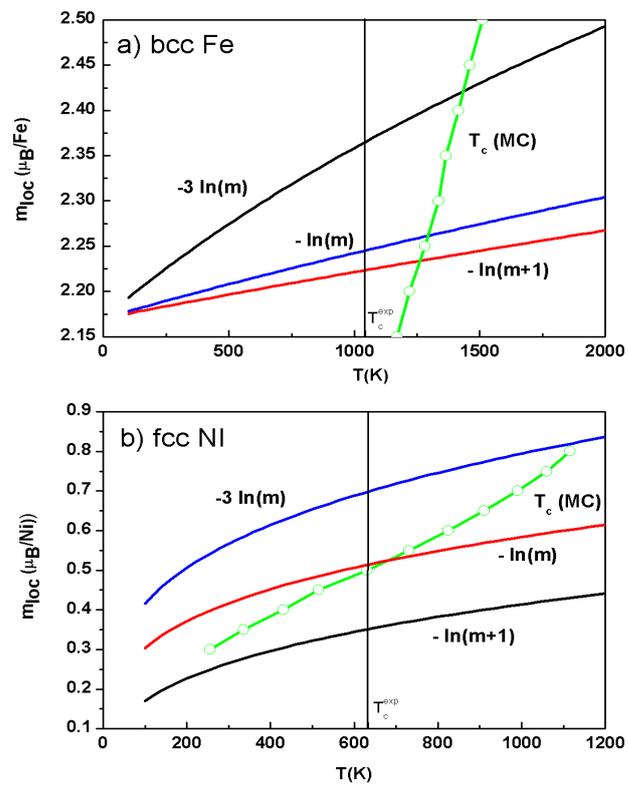

Figure 2_Khmelevskyi

# References.